\documentclass{article}
\usepackage{amsmath}
\usepackage{amsfonts}
\usepackage{amssymb}
\usepackage{amsthm}
\theoremstyle{definition}
\newtheorem{theorem}{Theorem}
\newtheorem{definition}{Definition}

\begin{document}
\title{Mathematical Foundations of Field Theory}
\author{Luther Rinehart\\ University of Pittsburgh\\ ldr22@pitt.edu}
\maketitle

\begin{abstract}
A mathematically rigorous Hamiltonian formulation for classical and quantum field theories is given. New results include clarifications of the structure of linear fields, and a plausible formulation for nonlinear fields.  Many mathematical formulations of field theory suffer greatly from either a failure to explicitly define the field configuration space, or else from the choice to define field operators as distributions.  A solution to such problems is given by instead using locally square-integrable functions, and by paying close attention to this space's topology.  One benefit of this is a clarification of the field multiplication problem: The pointwise product of fields is still not defined for all states, but it is densely defined, and this is shown to be sufficient for specifying dynamics.  Significant progress is also made, through this choice of configuration space, in appropriately representing field states with `infinitely many particles', or those which do not go to zero at infinity.
\end{abstract}

The purpose of this paper is to formulate a mathematical framework for quantum field theories.  In order to keep the construction as simple and as general as possible, a minimal amount of structure will be assumed, focusing only on what is essential to the concept of a quantum field.  Accordingly, I will consider a scalar field $\phi$ evolving in time on a spatial manifold $\Sigma$. The focus is on kinematics, rather than on constraining the particular form of the dynamics.  The formulation is presented in four steps: a linear classical field, a linear quantum field, a nonlinear classical field, and a nonlinear quantum field. Emphasis is on mathematical rigor.
\section{Linear Classical Field}
A linear field, also called a free field, is one for which the dynamical evolution is linear.  Its mathematical description is straightforward, and already well-understood.  

Let the manifold $\Sigma$ have measure $\mu$. The appropriate configuration space for a linear field is $\mathcal{F}\equiv L_2(\Sigma,\mu)$, the real Hilbert space consisting of the square integrable functions modulo sets of measure zero. From this configuration space, construct the phase space $\Phi\equiv\mathcal{F}\times\mathcal{F}^*$.  I will denote vectors in phase space by $\eta\equiv(\phi,\pi)$.

This space has a natural symplectic structure given by 
\begin{equation}
\Omega(\eta,\eta')=\pi'(\phi)-\pi(\phi').
\end{equation}
Dynamical evolution is given by a one-parameter subgroup of $\text{Sp}(\Phi,\Omega)$:
\begin{equation}
\eta(t)=U(t)\eta_0,
\end{equation}
where $U\in\text{Sp}(\Phi,\Omega)$, and 
\begin{equation}
U(t+s)=U(t)U(s).
\end{equation}
An illustrative example of how such a structure arises is as follows. Let $H$, the Hamiltonian, be any continuous quadratic function on $\Phi$:
\begin{equation}
H(\eta)=\frac{1}{2}H_{AB}\eta^A\eta^B,
\end{equation}
where $H_{AB}$ is a symmetric tensor on $\Phi$. Here and throughout, abstract index notation over $\Phi$ is employed whenever useful. Then if we take the equation of motion to be Hamilton's equation
\begin{equation}
\frac{d\eta^A}{dt}=\Omega^{AB}\nabla_BH,
\end{equation}
this is equivalent to
\begin{equation}\label{eqn:classical}
\frac{d\eta}{dt}=\hat{H}\eta,
\end{equation}
where $\hat{H}\equiv {H^A}_B$, and the symmetry of $H_{AB}$ is equivalent to $\hat{H}^\dag=-\hat{H}$, where the adjoint is taken with respect to $\Omega$.  Now integrating this equation of motion, we can identify
\begin{equation}
U=\exp (\hat{H}t),
\end{equation}
and the antihermiticity of $\hat{H}$ ensures that $U\in\text{Sp}(\Phi,\Omega)$.

Since $\Phi$ is infinite dimensional, not every linear map is continuous.  For most physical cases, $H$ is not continuous.  In the general case, Stone's theorem ensures that every one-parameter subgroup of $\text{Sp}(\Phi,\Omega)$ is generated by a densely defined, not necessarily continuous, anti-self-adjoint operator $\hat{H}$ (Ref. \cite{ref1}), so 
\begin{equation}
\eta(t)=\exp (\hat{H}t)\eta_0.
\end{equation}
The equation of motion is Hamilton's equation
\begin{equation}\label{eqn:Hamilton}
\frac{d\eta^A}{dt}=\Omega^{AB}\nabla_BH,
\end{equation}
where
\begin{equation}
H(\eta)=\frac{1}{2}\Omega(\eta,\hat{H}\eta),
\end{equation}
which may only be densely defined. Conversely, given a densely defined quadratic Hamiltonian, Hamilton's equation will generate a linear dynamical evolution on the domain of $H$, which extends uniquely to $\Phi$ as a one-parameter subgroup of $\text{Sp}(\Phi,\Omega)$.
\section{Linear Quantum Field}
\begin{definition}
(Ref. \cite{ref1}, p.4) Let $(\Phi,\Omega)$ be a complete symplectic vector space.  A \emph{Weyl system} over $\Phi$ is a continuous map $W:\Phi\rightarrow\text{U}(\mathbb{H})$, with $\mathbb{H}$ a complex Hilbert space, such that
\begin{equation}\label{eqn:ccr}
W(\eta)W(\eta')=\exp\left(-\frac{1}{2}i\Omega(\eta,\eta')\right)W(\eta+\eta').
\end{equation}
These are the canonical commutation relations.
\end{definition}
\begin{definition}
(Ref. \cite{ref1}, p.39) Let $G$ be a subgroup of $\text{Sp}(\Phi,\Omega)$. A $G$-covariant Weyl system over $\Phi$ is a Weyl system with a continuous unitary representation $\Gamma:G\rightarrow\text{U}(\mathbb{H})$ such that $\forall g\in G,\ \forall\eta\in\Phi$,
\begin{equation}
\Gamma(g)W(\eta)\Gamma(g)^{-1}=W(g\eta).
\end{equation}
\end{definition}
If $U$ represents the classical dynamical evolution, we want to require $U\in G$.  The linear quantum field should be a $U$-covariant Weyl system over $\Phi$.  The quantum state is a vector $\Psi\in\mathbb{H}$ satisfying $\langle\Psi,\Psi\rangle=1$, and it evolves in time according to
\begin{equation}\label{eqn:quantum}
\Psi(t)=\Gamma(U(t))\Psi_0.
\end{equation}
We now need to see how such a system is constructed. To begin, it is useful to promote $\Phi$ to a complex Hilbert space by giving it a complex structure.
\begin{definition}
A \emph{complex structure} on a real vector space is a  continuous linear map $J$ satisfying $J^2=-\mathbf{1}$.
\end{definition}
Given such a structure, the vector space becomes a vector space over $\mathbb{C}$, with
\begin{equation}
(a+ib)\eta\equiv a\eta+bJ\eta.
\end{equation}
\begin{definition}
A \emph{symplectic-compatible} complex structure on a symplectic vector space $\Phi$ is a complex structure $J$ satisfying \\
\\
1. $J\in\text{Sp}(\Phi,\Omega)$ \\
2. $\Omega(J\eta,\eta)\geq 0\ \forall\eta\in\Phi$
\end{definition}
Given such a structure, $\Phi$ acquires an inner product
\begin{equation}
\langle\eta,\eta'\rangle\equiv\Omega(J\eta,\eta')-i\Omega(\eta,\eta'),
\end{equation}
so if $\Phi$ is complete, it becomes a complex Hilbert space.
\begin{theorem}\label{thm:complex}
(Ref. \cite{ref1}, p.108) A complete symplectic vector space with linear dynamical evolution $U$ has at most one symplectic-compatible complex structure satisfying $[U,J]=0$.
\end{theorem}
The condition $[U,J]=0$ ensures that $U$ is unitary in the  new Hilbert space structure.  Consider the case in which the classical dynamical evolution is generated by a continuous anti-self-adjoint operator $\hat{H}$ (equation \ref{eqn:classical}).  Then the complex structure described in theorem \ref{thm:complex} exists and can be given explicitly, provided that $\hat{H}$ is positive definite and invertible. Then 
\begin{equation}
J=\frac{\hat{H}}{|\hat{H}|},
\end{equation}
where $|\hat{H}|$ is the positive square root of $\hat{H}^\dag \hat{H}$.
\\
An alternative method of obtaining a complex Hilbert space (Ref. \cite{ref2}) is to first form the complexification $\Phi^{\mathbb{C}}\equiv \Phi\times\Phi$. This space has natural complex structure 
\begin{equation}
J\equiv \left( \begin{matrix}
0 & -\mathbf{1} \\ 
\mathbf{1} & 0
\end{matrix} \right)
\end{equation}
and natural complex conjugation
\begin{equation}
C\equiv \left( \begin{matrix}
\mathbf{1} & 0 \\ 
0 & -\mathbf{1}
\end{matrix} \right).
\end{equation}
The symplectic form extends by complex linearity to 
\begin{equation}
\Omega^{\mathbb{C}} \equiv \left( \begin{matrix}
\Omega & i\Omega \\ 
i\Omega & -\Omega
\end{matrix} \right).
\end{equation}
Then the bilinear form on $\Phi^{\mathbb{C}}$ given by
\begin{equation}
\langle\eta,\eta' \rangle \equiv -i\Omega^{\mathbb{C}}(\bar{\eta},\eta')
\end{equation}
is non-degenerate and Hermitian.  We require a subspace $\Phi'$ of $\Phi^{\mathbb{C}}$ satisfying \\
1. $\langle,\rangle$ is positive definite on $\Phi'$ \\
2. $\Phi^{\mathbb{C}} = \Phi' \oplus \overline{\Phi'}$ \\
3. $\Phi'$ and $\overline{\Phi'}$ are orthogonal \\
Let $P$ be the projection operator onto $\Phi'$. For a dynamical evolution $U$, there is at most one such subspace satisfying $[U,P]=0$.  This construction is equivalent to choosing a symplectic-compatible complex structure $\tilde{J}$ on $\Phi$.  To see this, the projection operator $P$ is of the form
\begin{equation}
P=\frac{1}{2} \left( \begin{matrix}
\mathbf{1} & \tilde{J} \\ 
-\tilde{J} & \mathbf{1}
\end{matrix} \right),
\end{equation}
where $\tilde{J}$ is the required complex structure.  The associated inner product on $\Phi$
is $2\langle P\eta,P\eta' \rangle$. 
\\
The treatment of linear quantum fields will from here on be restricted to systems possessing complex structure as in theorem \ref{thm:complex}.  With this structure, $\Phi$ can be identified with the space of complex square-integrable functions on $\Sigma$.
\begin{definition}
Let $\Phi$ be a complex Hilbert space.  A \emph{linear quantum field} over $\Phi$ is a $\text{U}(\Phi)$-covariant Weyl system over $\Phi$ satisfying \\
\\
1. $\exists v\in\mathbb{H}$ such that $\Gamma(U)v=v\ \forall U\in\text{U}(\Phi)$, and Span$\{W(\eta)v\}$ is dense in $\mathbb{H}$. \\
2. For all positive self-adjoint $A$ on $\Phi$, the generator of $\Gamma(\exp(itA))$ is also positive.
\end{definition}
\begin{theorem}\label{thm:unitary}
(Ref. \cite{ref1}, p.64) All linear quantum fields over $\Phi$ are unitarily equivalent.
\end{theorem}
\begin{theorem}\label{thm:H-S}
(Ref. \cite{ref1}, p.133) Let $S\in\text{Sp}(\Phi,\Omega)$. $S$ is unitarily implementable on the linear quantum field over $(\Phi,\Omega,J)$ if and only if $[S,J]$ is Hilbert-Schmidt.
\end{theorem}
Also, if $J$ and $J'$ are symplectic-compatible complex structures on $(\Phi,\Omega)$, then the linear quantum fields over $(\Phi,\Omega,J)$ and $(\Phi,\Omega,J')$ are unitarily equivalent if and only if the following are satisfied:\\
1. $\exists S\in\text{Sp}(\Phi,\Omega)$ such that $J'=SJS^{-1}$ \\
2. $J-J'$ is Hilbert-Schmidt. \\
The linear quantum field over a complex Hilbert space $\Phi$ can be explicitly constructed via the `particle representation.' Define the $n$-particle Hilbert space to be the symmetric $n$-tensor product of $\Phi$:
\begin{equation}
\Phi^n\equiv\text{Sym}\left(\underset{n}{\underbrace{\Phi\otimes...\otimes\Phi}}\right), \qquad \Phi^0\equiv\mathbb{C},
\end{equation}
and define Fock space
\begin{equation}
\mathbb{H}\equiv\bigoplus^\infty_{n=0} \Phi^n,
\end{equation}
understood as the natural Hilbert space direct sum.  Write $\Psi\equiv\bigoplus^\infty_{n=0} \Psi_n$ for the state.  Now for each $\eta\in\Phi$, associate the following densely defined, discontinuous linear operator on $\mathbb{H}$ (Ref. \cite{ref2}, appendix):
\begin{equation}
a_\eta(\Psi)\equiv\bigoplus^\infty_{n=0}\left(\sqrt{n}\langle\eta,\Psi_n\rangle\right),
\end{equation}
where $\langle\eta,\Psi_n\rangle$ is understood to mean $\eta^\dag_{A_1}\Psi^{A_1...A_n}_n$ with indices over $\Phi$. $a_\eta$ has adjoint given by
\begin{equation}
a^\dag_\eta(\Psi)\equiv\bigoplus^\infty_{n=0}\left(\sqrt{n+1}\ \text{Sym}(\eta\otimes\Psi_n)\right).
\end{equation}
\begin{theorem}
These operators satisfy
\begin{subequations}
\begin{equation}
[a_\eta,a_\psi]=[a^\dag_\eta,a^\dag_\psi]=0, 
\end{equation}
\begin{equation}
[a_\eta,a^\dag_\psi]=\langle\eta,\psi\rangle\mathbf{1}.
\end{equation}
\end{subequations}
\end{theorem}
\begin{definition}
The \emph{vacuum} $v\equiv(1,0,0,0,...)\in\mathbb{H}$, which satisfies
\begin{equation}
a^\dag_\eta v=\eta\qquad a_\eta v=0.
\end{equation}
\end{definition}
\begin{definition}
Let $A\equiv\{a^\dag_\eta | \eta\in\Phi \}$, and $A^*\equiv\{a_\eta | \eta\in\Phi \}$.
\end{definition}
\begin{theorem}
\begin{subequations}
\begin{equation}
a_{(\alpha\eta+\beta\psi)}=\bar{\alpha}a_\eta+\bar{\beta}a_\psi,
\end{equation}
\begin{equation}
a^\dag_{(\alpha\eta+\beta\psi)}=\alpha a^\dag_\eta+\beta a^\dag_\psi.
\end{equation}
\end{subequations}
\end{theorem}
So $A$ and $A^*$ are vector spaces, and we can see that there is a natural isometric isomorphism between $A$ and $\Phi$, and between $A^*$ and $\Phi^*$.
\begin{definition}
Let $\{\eta_\mu \}$ be an orthonormal basis for $\Phi$. The \emph{field operator} is the element of $\Phi\otimes A^*$ given by 
\begin{equation}
\hat{\psi}^A\equiv\sum\eta^A_\mu\otimes a^\mu.
\end{equation}
The vector index is over $\Phi$, and $a^\mu$ corresponds to $\eta_\mu$.  This is independent of basis, and it is clear that the sum converges in the natural topology on $\Phi\otimes A^*$.  We also have 
\begin{equation}
\hat{\psi}^\dag_A\equiv\sum{\eta_A^\mu}^\dag\otimes a^\dag_\mu.
\end{equation}
\end{definition}
Note that these are not defined pointwise on $\Sigma$ for the same reason that square-integrable functions are not defined pointwise.
\begin{flushleft}
\textbf{Useful Identities}
\end{flushleft}
\begin{equation}
\begin{split}
&[\hat{\psi}^A, \hat{\psi}^B]=0 \\
&[\hat{\psi}^A, \hat{\psi}^\dag_B]=\delta^A_{\ B} \otimes \mathbf{1} \\
&\eta^\dag_A\hat{\psi}^A = a_\eta  \\
&\eta^A\hat{\psi}^\dag_A = a^\dag_\eta \\
&[\hat{\psi}^A, a^\dag_\eta] = \eta^A\otimes \mathbf{1}
\end{split}
\end{equation}
\begin{definition}
The \emph{number operator} for $\eta$ is $\hat{n}_\eta\equiv a^\dag_\eta a_\eta$. The \emph{total number operator} is $\hat{N}\equiv \hat{\psi}^\dag_A\hat{\psi}^A = \sum a^\dag_\mu a^\mu$.

Let ${T^A}_B$ be any densely defined self-adjoint operator on $\Phi$. Define the \emph{total $T$ operator} on $\mathbb{H}$ as $\hat{T}\equiv \hat{\psi}^\dag_A {T^A}_B \hat{\psi}^B$.

Finally, since the pointwise product of two $L_2$ functions is naturally in $L_1$, we can loosely define density operators
\begin{equation}
\hat{n}(x)\equiv \hat{\psi}^\dag(x)\hat{\psi}(x),
\end{equation}
\begin{equation}
\hat{t}(x)\equiv \text{Re}\left( \hat{\psi}^\dag(x)T(\hat{\psi})(x)\right).
\end{equation}
\end{definition}
It remains to show how this construction realizes the linear quantum field over $\Phi$.  For $\eta\in\Phi$ define 
\begin{subequations}
\begin{equation}
\hat{\phi}(\eta)\equiv \frac{1}{\sqrt{2}}(a_\eta+a^\dag_\eta),
\end{equation}
\begin{equation}
W(\eta)\equiv \exp i\hat{\phi}(\eta).
\end{equation}
\end{subequations}
Since $\hat{\phi}(\eta)$ is densely defined self-adjoint, $W(\eta)$ extends uniquely to a unitary operator on $\mathbb{H}$.  Additionally, for $U\in\text{U}(\Phi)$, define $\Gamma(U)$ to be the natural tensor product - direct sum action of $U$ on $\mathbb{H}$, acting as identity on $\Phi^0$.
\begin{theorem}
(Ref. \cite{ref1}, p.49) $(W,\Gamma,v)$ is a linear quantum field over $\Phi$.
\end{theorem}
With the added complex structure on $\Phi$, the classical equation of motion (equation \ref{eqn:classical}) can be written as 
\begin{equation}
i\frac{d\eta^A}{dt}={H^A}_B \eta^B.
\end{equation}
where ${H^A}_B$ is densely defined self-adjoint.  It can be shown that the quantum dynamics (equation \ref{eqn:quantum}) can be described by the equation of motion 
\begin{equation}
i\frac{d\Psi}{dt}=\hat{H}\Psi,
\end{equation}
where 
\begin{equation}
\hat{H}=\sum a^\dag_\mu {H^\mu}_\nu a^\nu=\hat{\psi}^\dag_A {H^A}_B \hat{\psi}^B.
\end{equation}
The linear quantum field has certain shortcomings, most obviously its restriction to linear dynamical evolution.  Theorem \ref{thm:unitary} on unitary equivalence only applies to U$(\Phi)$, so quantum representations of arbitrary symplectomorphisms will not necessarily be unitarily equivalent, either to each other, or to the representation given above for the linear field.  This is a statement to the effect that the Stone-von Neumann theorem does not hold in infinite dimensions.  Indeed, if $\Phi$ is finite-dimensional, then all symplectic transformations are unitarily representable, and all Weyl systems are unitarily equivalent. But, as theorem \ref{thm:H-S} shows, if $\Phi$ is infinite-dimensional then this fails (there can be non-Hilbert-Schmidt operators in infinite dimensions). There are then many unitarily inequivalent representations of the canonical commutation relations (equation \ref{eqn:ccr}).  This seems to be the main mathematical issue facing quantum field theory.  In the case of a free field, it is overcome by the construction described above, in which the evolution selects a preferred representation, but for an interacting theory the issue is inescapable. Consequently, as will be seen later, the appropriate context for doing quantum field theory in a representation-independent way is not with Hilbert spaces, but rather with C*-algebras.

The other shortcoming of the linear quantum field is the restriction to square-integrable fields.  This gives convenient algebraic and topological structure, but it cannot be a description of nature, since most physically realizable field states in nature do not go to zero at infinity, such as in a homogeneous, non-compact universe containing infinitely many particles.  Related to this problem is the fact that the Fock space representation has only states with arbitrarily large but still finite numbers of particles. There are no states with infinitely many particles.  This is a subtle point, because it is technically possible to make a superposition of states of finitely many particles such that the expectation value of the total number operator diverges.  To name the issue more precisely, the eigenstates of the total number operator form a complete basis, so all states are superpositions of those with finitely many particles. A state that properly has infinitely many particles ought to be orthogonal to all of these.  
\section{Nonlinear Classical Field}
The appropriate configuration space for nonlinear fields is $\mathcal{F}\equiv L^{loc}_2(\Sigma,\mu)$, the \textit{locally} square-integrable functions, which are the functions (modulo sets of measure zero) which are square integrable on all compact subsets of $\Sigma$.  These functions are well suited to describing physical fields states, since they have the same local behavior as in the linear case, plus the ability to describe arbitrary behavior at infinity.

$\mathcal{F}$ is not a Hilbert space; it is not even normable.  However,
\begin{theorem}
$\mathcal{F}$ is a complete metrizable space.  In fact it is a Frechet space (Ref. \cite{ref3}).
\begin{proof}
The natural topology on $\mathcal{F}$ is that in which a sequence of functions converges if and only if its restriction to every compact subset $U\subseteq\Sigma$ converges in $L_2(U)$. Let $\{U_n\}$ be a countable partition of $\Sigma$ into subsets with compact closure.  Then $\mathcal{F}=\prod L_2(U_n)$ with the product topology.  Let $\|\cdot\|_n$ be the natural norm on $L_2(U_n)$. For $\phi,\phi'\in\mathcal{F}$, define
\begin{equation}
d(\phi,\phi')\equiv\sum\frac{1}{2^n}\frac{\|\phi-\phi'\|_n}{1+\|\phi-\phi'\|_n}.
\end{equation}
Then $d$ is a metric on $\mathcal{F}$, and $\mathcal{F}$ is Frechet (Ref. \cite{ref3}, p.40).
\end{proof}
\end{theorem}
\begin{theorem}
The continuous dual space of $\mathcal{F}$ is $\mathcal{F}^*=L^{comp}_2(\Sigma,\mu)$, the square integrable functions of compact support. $\mathcal{F}^*$ is dense in $\mathcal{F}$ under the natural embedding.
\begin{proof}
To show that $L^{comp}_2$ is dense in $\mathcal{F}$, again let $\{U_n\}$ be a countable partition of $\Sigma$ into subsets with compact closure.  Let $\{e_{nm}\}$ be an orthonormal basis for $L_2(U_n)$, so that $\{e_{nm}\}$ is an orthonormal Schauder basis for $\mathcal{F}$. It is clear that every finite linear combination of the $\{e_{nm}\}$ will have compact support, so Span$\{e_{nm}\}\subseteq L^{comp}_2$. Since $\{e_{nm}\}$ is a Schauder basis, its span is dense, so it follows that $L^{comp}_2$ is dense.

To show that $L^{comp}_2\subseteq \mathcal{F}^*$, it is clear that every compactly-supported square-integrable function gives a linear functional on $\mathcal{F}$ under the pairing
\begin{equation}
\psi(\phi)=\int_{\Sigma} \psi\phi\ d\mu,
\end{equation}
and it is continuous, as follows from its continuity on $L_2$.

Finally, to show that $\mathcal{F}^*\subseteq L^{comp}_2$, an arbitrary continuous linear functional $\psi$ can be characterized by its action on the Schauder basis $\{e_{nm}\}$. Suppose $\psi(e_{nm})\neq 0$ for infinitely many $n$. Then since $\mathcal{F}=\prod L_2(U_n)$, one can choose ${\alpha_n}$ such that $\sum\alpha_n \psi(e_{nm})$ does not converge, which gives a contradiction.  Thus $\psi(e_{nm})\neq 0$ for only finitely many $n$. Since $\psi$ is continuous, for a given $n$, $\sum e_{nm}\psi(e_{nm})$ converges to a function in $L_2(U_n)$. Then $\sum\sum e_{nm} \psi(e_{nm})$ gives a function in $L^{comp}_2$. By orthonormality,
\begin{equation}
\psi(\phi)=\int_{\Sigma} \sum\sum e_{nm} \psi(e_{nm})\phi\ d\mu,
\end{equation}
\end{proof}
\end{theorem}
Another important property is that $\mathcal{F}$ is reflexive. That is, $\mathcal{F}^{**}=\mathcal{F}$ (Ref. \cite{ref4}). The three spaces $\mathcal{F}^*\subseteq L_2 \subseteq \mathcal{F}$ form a Gelfand triple.

We know from experience that we need a pair of real-valued fields to specify boundary conditions, so take as phase space $\Phi\equiv\mathcal{F}\times\mathcal{F}$. I will denote vectors in both phase space and dual phase space by $\eta\equiv(\phi,\pi)$.

This is not a symplectic vector space, but it does have a Poisson structure. First define the bilinear map $\Omega:\Phi^*\times\Phi^*\rightarrow\mathbb{R}$,
\begin{equation}
\Omega(\eta,\eta')\equiv\langle\phi,\pi'\rangle-\langle\pi,\phi'\rangle,
\end{equation}
where the inner products are in $L_2$. 
\begin{definition}
The \emph{Poisson bracket} of smooth, complex-valued functions is $\{\cdot\}:C^\infty(\Phi)\times C^\infty(\Phi)\rightarrow C^\infty(\Phi)$
\begin{equation}
\{f,g\}\equiv \Omega(\nabla f,\nabla g),
\end{equation}
where the derivative on $\Phi$ is the Frechet derivative.  
\end{definition}
$\Omega$ is a symplectic structure on $\Phi^*$, and it gives a continuous injective linear map $\omega:\Phi^*\rightarrow \Phi$, which takes $\eta$ to the unique vector satisfying $\forall \psi\in\Phi^*$,
\begin{equation}
\psi(\omega(\eta))=\Omega(\psi,\eta).
\end{equation}
Explicitly,
\begin{equation}
\omega\left( (\phi,\pi)\right)=(\pi,-\phi).
\end{equation}
Dynamical evolution is a one-parameter group of symplectomorphisms of $\Phi$:
\begin{equation}
\eta(t)=U(\eta_0,t),
\end{equation}
where $U(t):\Phi\rightarrow \Phi$ is continuous, preserves the Poisson bracket , and satisfies
\begin{equation}
U(t+s)=U(t)\circ U(s).
\end{equation}
This dynamical evolution can be generated by a densely defined, smooth, real-valued function $H$ on $\Phi$, according to the equation of motion
\begin{equation}
\frac{d\eta}{dt}=\omega(\nabla H).
\end{equation}
$H$ need only be densely defined on $\Phi$, since the full evolution can be extracted by extension, just as in the linear case (equation \ref{eqn:Hamilton}).  Not every such Hamiltonian $H$ will generate a well-defined one-parameter group of symplectomorphisms. Some may lead to singular solutions of various sorts.  Therefore, we restrict attention to those Hamiltonians that do generate one-parameter groups of symplectomorphisms.\\
A note on multiplication of fields: neither $L_2$ nor $L^{loc}_2$ is an algebra under pointwise multiplication.  However, in both cases, pointwise multiplication by a vector $\eta$ is a densely defined linear operator (which is continuous and everywhere-defined if $\eta$ happens to be bounded). Thus it is sensible and well-defined to consider Hamiltonians that include pointwise products of fields.  This issue is not the origin of divergence in the quantized theory.  That is due to the existence of inequivalent representations, so that the representation of $\hat{H}$ fails to be densely defined on the Hilbert space $\mathbb{H}$, and so fails to generate a well-defined representation of the evolution.
\section{Nonlinear Quantum Field}
The solution to the problem of unitarily inequivalent representations is to realize quantum mechanics in the more general setting of C*-algebras (Ref. \cite{aqft}). In this approach, the algebra of observables plays the fundamental physical role, instead of a Hilbert space of states.  Let $\mathbb{A}$ be a C*-algebra. The quantum state is a positive linear functional $E$ on $\mathbb{A}$, satisfying $E(1)=1$. The dynamical evolution is a one-parameter group of automorphisms of $\mathbb{A}$. 

By analogy with classical mechanics, a first possibility for the choice of $\mathbb{A}$ is the space of bounded, $C^\infty$ complex functions on $\Phi$, where $\Phi$ is as for nonlinear classical fields. This space has a natural norm $\|f\|\equiv \sup |f(\eta)|$. However, it will be easier for the following constructions to work first with the smaller space $\mathbb{A}_0$ defined as follows: define the function $W:\Phi^*\rightarrow C^\infty(\Phi)$ as $W(\eta)(\psi)\equiv \exp i\eta(\psi)$, and then take $\mathbb{A}_0\equiv \text{Span}\{W(\eta) | \eta\in\Phi^* \}$.
\begin{definition}
The \emph{star product} (Ref. \cite{ref5}) on $C^\infty(\Phi)$ is 
\begin{equation}
\begin{split}
f\star g &\equiv \sum_{n=0}^{\infty} \frac{1}{n!}\left( \frac{i}{2} \right)^n \nabla_{A_1}\cdots \nabla_{A_n} f\  \Omega^{A_1B_1}\cdots\Omega^{A_nB_n}\ \nabla_{B_1}\cdots\nabla_{B_n} g 
\\
\\
&= f \exp \left(\frac{1}{2}i\Omega(\overleftarrow{\nabla},\overrightarrow{\nabla}) \right) g.
\end{split}
\end{equation}
\end{definition}
\begin{theorem}
$\mathbb{A}_0$ with $\star$, complex conjugation, and supremum norm, is a normed *-algebra with 1.
\end{theorem}
The following theorem states that the canonical commutation relations are satisfied.
\begin{theorem}
\begin{equation}
W(\eta)\star W(\eta')=\exp\left(-\frac{1}{2}i\Omega(\eta,\eta')\right)W(\eta+\eta').
\end{equation}
\end{theorem}
\begin{definition}
Let $\mathbb{A}$ be the enveloping C*-algebra of $\mathbb{A}_0$ (Ref. \cite{ref6}, p.47; Ref. \cite{ref7}, p.151).
\end{definition}
\begin{definition}
Let $\hat{\Phi}^*$ denote the space $\Phi^*$ understood as a subset of $C^\infty(\Phi)$ with the star product.  Denote elements of $\hat{\Phi}^*$ as $\hat{\eta}$.
\end{definition}
\begin{theorem}
\begin{equation}
[\hat{\eta},\hat{\eta}']=i\Omega(\eta,\eta').
\end{equation}
\end{theorem}
\begin{definition}
Let $\{\eta_\mu\}$ be a compactly-supported orthonormal basis for $\Phi$, so that the coordinate functionals $\{\eta^\mu\}$ are an orthonormal basis for $\Phi^*$.  The \emph{field operator} is the element of $\Phi\otimes\hat{\Phi}^*$ given by 
\begin{equation}
\hat{\psi}^A \equiv \sum \eta^A_\mu \otimes \hat{\eta}^\mu.
\end{equation}
\end{definition}
The vector index is over $\Phi$. This is independent of basis, and it is clear that the sum converges in the natural topology on $\Phi\otimes\hat{\Phi}^*$.
\begin{flushleft}
\textbf{Useful Identities}
\end{flushleft}
\begin{equation}
\begin{split}
&[\hat{\psi}^A,\hat{\psi}^B] = i\Omega^{AB}\otimes\mathbf{1} \\
&\eta_A \hat{\psi}^A = \hat{\eta} \\
&[\hat{\eta},\hat{\psi}^B] = i\eta_A \Omega^{AB} \otimes\mathbf{1} \\
\end{split}
\end{equation}
Since these field operators are not elements of $\mathbb{A}$, we need to define how a quantum state $E$ acts on them.  
\begin{definition}
$E(W(\eta))$ gives a function on $\Phi^*$.  Define the \emph{expectation value} of the field operator to be the element of $\Phi$ given by
\begin{equation}
\langle\hat{\psi}^A\rangle \equiv -i\nabla^A E(W(\eta))|_{\eta=0},
\end{equation}
where again, the derivative is the Frechet derivative. Higher $n$-point functions can be defined similarly.
\end{definition}
\begin{theorem}\label{thm:rep}
All linear symplectomorphisms of $\Phi$ can be represented by *-automorphisms of $\mathbb{A}$, that is, there is a unique representation $\Gamma: \text{Sp}(\Phi)\rightarrow \text{Aut}(\mathbb{A})$ satisfying
\begin{equation}\label{eq:rep}
\Gamma(UU')=\Gamma(U)\Gamma(U'),
\end{equation}
and $\forall U\in\text{Sp}(\Phi),\ \forall\eta\in\Phi^*$,
\begin{equation}\label{eq:covariant}
\Gamma(U)\left(W(\eta)\right)=W(U(\eta)).
\end{equation}
\begin{proof} 
The action of $\Gamma(U)$ on $W(\eta)$ is given. It is clear that it satisfies equation \ref{eq:rep}.  The $\{W(\eta)\}$ are linearly independent.  Extend the action of $\Gamma(U)$ by linearity to $\mathbb{A}_0$, that is,
\begin{equation}
\Gamma(U)\left(\sum\alpha_i W(\eta_i) \right)=\sum\alpha_i \Gamma(U)\left(W(\eta_i) \right).
\end{equation}
This gives a continuous linear isomorphism of $\mathbb{A}_0$, which extends uniquely to a continuous linear isomorphism of $\mathbb{A}$.  It is straightforward to show that 
\begin{equation}
\Gamma(U)(f^*)=\left(\Gamma(U)(f) \right)^*,
\end{equation}
and the automorphism property
\begin{equation}
\Gamma(U)(f\star g) =\Gamma(U)(f)\star \Gamma(U)(g)
\end{equation}
follows from the symplectic property of $U$:
\begin{equation}\label{eq:automorphism}
\begin{split}
\Gamma(U)\left(W(\eta)\right)\star \Gamma(U)\left(W(\eta')\right)&=W(U(\eta))\star W(U(\eta')) \\
&= \exp\left(-\frac{1}{2}i\Omega(U(\eta),U(\eta'))\right)W(U(\eta)+U(\eta')) \\
&= \exp\left(-\frac{1}{2}i\Omega(\eta,\eta')\right)W(U(\eta+\eta')) \\
&= \Gamma(U)\left(W(\eta)\star W(\eta')\right).
\end{split}
\end{equation}
So $\Gamma(U)$ is a *-automorphism of $\mathbb{A}$.
\end{proof}
\end{theorem}
Theorem \ref{thm:rep} is useful for implementing linear symmetry transformations, but it does not hold for nonlinear symplectomorphisms.  The above proof explicitly uses linearity in equation \ref{eq:automorphism}, and this shows that if $U$ is nonlinear, then there does not exist an automorphism $\Gamma(U)$ satisfying equation \ref{eq:covariant}.

To obtain a one-parameter group of automorphisms from a classical nonlinear evolution generated by $H$, a first guess is, for $f\in\mathbb{A}$, to take the equation of motion
\begin{equation}
i\frac{df}{dt}=H\star f - f\star H.
\end{equation}
However, this is only densely defined in $\Phi$, so it is not clear that it gives a well-defined mapping on $\mathbb{A}$. Instead, if $U$ is the diffeomorphism of $\Phi$ giving the classical evolution, transform the functions $f$ in $\mathbb{A}$ exactly as they do classically:
\begin{equation}
f'=f\circ U,
\end{equation}
which is just the natural diffeomorphism action on functions.  This is a linear isomorphism of $\mathbb{A}$ as a vector space. 
\begin{theorem}
\begin{equation}
(f\circ U)\star(g\circ U)=(f\star g)\circ U.
\end{equation}
\begin{proof}
Let $DU$ denote the derivative of $U$. $U$ being a symplectomorphism is equivalent  to
\begin{equation}
{DU^A}_B {DU^C}_D \Omega^{BD}=\Omega^{AC}.
\end{equation}
Now we have
\begin{equation}
\begin{split}
&(f\circ U)\star (g\circ U) =
\\
&=\sum_{n=0}^{\infty} \frac{1}{n!}\left( \frac{i}{2} \right)^n \nabla_{A_1}\cdots \nabla_{A_n} (f\circ U)\  \Omega^{A_1B_1}\cdots\Omega^{A_nB_n}\ \nabla_{B_1}\cdots\nabla_{B_n} (g\circ U) 
\\
\\
&= \bigg{[} \sum_{n=0}^{\infty} \frac{1}{n!}\left( \frac{i}{2} \right)^n \nabla_{A_1}\cdots \nabla_{A_n} f\ {DU^{A_1}}_{C_1}\cdots{DU^{A_n}}_{C_n}\ \Omega^{C_1D_1}\cdots\Omega^{C_nD_n}
\\
&\qquad\qquad\qquad\qquad\nabla_{B_1}\cdots\nabla_{B_n} g\ {DU^{B_1}}_{D_1}\cdots{DU^{B_n}}_{D_n}\bigg{]} \circ U
\\
\\
&=(f\star g)\circ U.
\end{split}
\end{equation}
\end{proof}
\end{theorem}
This means we have an automorphism of $\mathbb{A}$. In the case of a linear symplectomorphism, it agrees with the construction given in theorem \ref{thm:rep}.

\section{Covariant formulation}
The problem of finding a one-parameter group of automorphisms to give dynamical evolution can also be approached using the covariant formulation (Ref. \cite{covariant}, Ref. \cite{ref2}). The strategy is to take phase space $\Phi$ to be the space of all solutions to the classical equations of motion, based on the idea that the space of initial data should be isomorphic to the space of solutions.  We change perspective from a field $\phi$ evolving in time on a manifold $\Sigma$, to a static field solution $\phi$ on a larger manifold $M$, which at least locally looks like $\Sigma\times\mathbb{R}$, the $\mathbb{R}$ representing the time evolution. To be exact, we consider the space of all smooth solutions to some classical field equation for $\phi$ on $M$, and then take phase space $\Phi$ to be the closure of this space in $L^{loc}_2(M)$.  If the theory is nonlinear, then in principle $\Phi$ is not a vector space, but an infinite dimensional manifold.\\
The tangent space to the manifold $\Phi$ at a particular solution $\phi_0$ is the set of (locally square-integrable) solutions to the linearized field equation around $\phi_0$. Let $L_{\phi_0}$ denote this linear equation. The cotangent space $T\Phi^*_{\phi_0}$ has a natural symplectic form, given by 
\begin{equation}
\Omega=R-A,
\end{equation}
where $R$ and $A$ are respectively the retarded and advanced propagators of $L_{\phi_0}$.  This can equivalently be expressed as
\begin{equation}
\Omega(\phi_1,\phi_2)= \int_{\Sigma} (\phi_1 \Pi^\mu_2 - \phi_2 \Pi^\mu_1) n_\mu,
\end{equation}
a surface integral over an initial-data surface $\Sigma$ with normal $n$, where $\Pi$ is the canonical momentum vector associated with $L_{\phi_0}$. This expression is independent of the choice of $\Sigma$, because the vector expression being integrated is divergence-free by virtue of the equation $L_{\phi_0}$ satisfied by $\phi_1$ and $\phi_2$.\\
With this structure, the covariant phase space $\Phi$ becomes a Poisson manifold.  Also, if the space of initial conditions is linear, there will be a canonical flat derivative operator $\nabla$ on $\Phi$, so that $\Phi$ has sufficient structure to define a star-product on $C^\infty(\Phi)$: 
\begin{equation}
\begin{split}
f\star g &\equiv \sum_{n=0}^{\infty} \frac{1}{n!}\left( \frac{i}{2} \right)^n \nabla_{A_1}\cdots \nabla_{A_n} f\  \Omega^{A_1B_1}\cdots\Omega^{A_nB_n}\ \nabla_{B_1}\cdots\nabla_{B_n} g 
\\
\\
&= f \exp \left(\frac{1}{2}i\Omega(\overleftarrow{\nabla},\overrightarrow{\nabla}) \right) g.
\end{split}
\end{equation}
The construction of the C*-algebra and the quantum theory over $\Phi$ then proceeds as before, only now dynamical evolution is already included in the structure of the algebra.

Unfortunately, this formulation does not work for many of the interacting field theories commonly studied, because they include nonlinear initial-value constraints, so the space of initial conditions does not form a vector space.  Such theories would require a theory of quantization of an arbitrary Poisson manifold.

\section{Note on Distributions}
This paper takes the position that it is more appropriate to use $L_2$ and $L_2^{loc}$ as field configuration space, rather than the space of distributions. The justification is that this choice leads to a satisfactory mathematical formulation. However, because this position is contrary to convention, it is appropriate to explain further why distributions need not (should not) be used. On an aesthetic level, the space of distributions is so extremely large and has a topology so obscure that it is unlikely to have any relevance to describing physical phenomena. But for more physical justification, the following is a list of reasons sometimes given for using distributions, along with why each is unconvincing.\\
1. \emph{We need to use distributions to describe fields with arbitrary (non-square-integrable) behavior at infinity.} This is a very desirable characteristic of the field configuration space, and it is solved by using $L_2^{loc}$.\\
2. \emph{We need to be able to use the Dirac $\delta$-function (such as in field commutation relations), and this is a distribution.}  In the context in which the Dirac $\delta$-function is used in field theory, its correct mathematical interpretation is as the identity linear map on $L_2$.\\
3. \emph{The mode-sum definition of the field operator does not converge pointwise, and so the field operator cannot be defined pointwise and must be a distribution.}  The first part is true, but this non-pointwise-defined property is also characteristic of $L_2$ and $L_2^{loc}$. The mode-sum converges in $L_2$ and $L_2^{loc}$.\\
4. \emph{We need fields to be infinitely differentiable.} This is not quite correct. The derivative is a densely-defined operator on $L_2$ and $L_2^{loc}$, and as discussed above, a Hamiltonian need only be densely defined to give a well-defined dynamical evolution. A similar situation applies to the pointwise multiplication issue.  This is exactly analogous to single-particle quantum mechanics, in which everyone agrees that the state space is $L_2$, but we freely employ derivative operators that are only densely-defined.

\section{Discussion}
A well known symptom of a lack of mathematical rigor in quantum field theory is the presence of `divergences'.  These divergences are not due to any physical failure of the theory, but rather they result from faulty mathematical assumptions, typically by assuming the existence of some operator which does not, in fact, exist. The mathematical reason for most of the divergences in quantum field theory is the failure of the Stone-von Neumann theorem in infinite dimensions: not all Hilbert-space representations of the canonical commutation algebra are unitarily equivalent (This happens to be one way of expressing a result known as Haag's theorem (Ref. \cite{ref8})). As a corollary, not all classical nonlinear time evolutions can be represented with unitary maps.  We get a divergence when attempting to calculate matrix elements of the non-existent unitary map. (Ref. \cite{ref1} chapter 4) \\
Because of the unitary inequivalence in infinite dimensions, formulating a Hilbert space representation involves a nontrivial choice. The choice of representation can be expressed many different ways, including as a choice of complex structure on phase space as presented above, or as a choice of invariant inner product on phase space, or the choice of a vacuum state on which to build the theory, or a choice of operator-ordering.  Having made such a choice, there is then no reason to expect all physically relevant  states to lie in an equivalent representation (Ref. \cite{ref2}, Ref. \cite{ref8}).  An example of this is the failure of the Fock representation built off of the free vacuum to include field states which are not square-integrable, or which have infinitely many particles.  This might be related to infrared divergences.\\
These failures suggests that nature should not be described by a Hilbert space representation, but rather directly by the abstract algebra of observables, as discussed above.  This approach, combined with the use of \textit{locally} square-integrable functions to capture infrared behavior, can help to resolve the barriers to formulating a mathematically well-defined theory.

\end{document}